\documentclass[10pt,journal,doublecolumn]{IEEEtran}
\usepackage{amsmath,amssymb,threeparttable,placeins,enumerate,caption}
\usepackage{mathtools,relsize,cite}
\usepackage[pdftex]{graphicx}
\usepackage{epstopdf}
\usepackage[usenames]{color}
\usepackage{amsfonts}
\usepackage{latexsym}
\usepackage{subfigure,multirow,makecell,stfloats}

\definecolor{burntorange}{rgb}{0.8, 0.33, 0.0}

\usepackage{algorithm}
\usepackage[noend]{algpseudocode}

\newtheorem{theorem}{{{\textit{Theorem}}}}

\newtheorem{definition}{{{\textit{Definition}}}}

\newtheorem{remark}{{{\textit{Remark}}}}

\newtheorem{example}{{{\textit{Example}}}}

\begin{document}
	
	
	\title{Generalized Constructions of Complementary Sets of Sequences of Lengths Non-Power-of-Two}
	\author{Gaoxiang Wang,~Avik Ranjan Adhikary,~Zhengchun Zhou,~Yang Yang

		\thanks{
			Manuscript Submitted 31st October, 2019.
			
			
			Gaoxiang Wang is with Department of Mathematics, Southwest Jiaotong University, China, e-mail: {\tt 979338034@qq.com}. Avik R. Adhikary is with the School of Information Science and Technology, Southwest Jiaotong University, China, e-mail: {\tt avik.adhikary@ieee.org}. Zhengchun Zhou and Yang Yang are with Department of Mathematics, Southwest Jiaotong University, China, e-mail: {\tt zzc@swjtu.edu.cn;~yang\_data@qq.com}.
		 }
	}
	\maketitle
	
	\begin{abstract}
		The construction of complementary sets (CSs) of sequences with different set size and sequence length become important due to its practical application for OFDM systems. Most of the constructions of CSs, based on generalized Boolean functions (GBFs), are of length $2^\alpha$ ($\alpha$ is a natural number). Recently some works have been reported on construction of CSs having lengths non-power of two, i.e., in the form of $2^{m-1}+2^v$ ($m$ is natural number, $0\leq v <m $), $N+1$ and $N+2$, where $N$ is a length for which $q$-ary complementary pairs exist. In this paper, we propose a construction of CSs of lengths $M+N$ for set size $4n$, using concatenation of CSs of lengths $M$ and $N$, and set size $4n$, where $M$ and $N$ are lengths for which $q$-ary complementary pairs exists. Also, we construct CSs of length $M+P$ for set size $8n$ by concatenating CSs of lengths $M$ and $P$, and set size $8n$, where $M$ and $P$ are lengths for which $q$-ary complementary pairs and complementary sets of size $4$ exists, respectively. The proposed constructions cover all the previous constructions as special cases in terms of lengths and lead to more CSs of new sequence lengths which have not been reported before.  
	\end{abstract}

\vspace{-0.2cm}
	
	\section{Introduction}
	\IEEEPARstart{I}{n} 1950, M. J. Golay introduced complementary pairs in his work on multislit spectrometry \cite{golay1}. Golay complementary pairs (GCPs) are pair of sequences whose aperiodic autocorrelation sums (AACSs) are zero everywhere, except at the zero shift \cite{golay2}. Binary GCPs are available only for limited lengths of the form $2^\alpha10^\beta26^\gamma$ (where $\alpha$, $\beta$, and $\gamma$ are non-negative integers)\cite{golay2,Borwein03}. In 1972, Tseng and Liu \cite{tseng72} extended the idea of complementary pairs to complementary sets (CSs) of sequences. CSs found a number of applications in communication systems \cite{spasojevic,davis99,paterson,popovic,schmidt,palash1}.
	
	Modern day communication systems, such as orthogonal frequency division multiplexing (OFDM) systems, often suffers from high peak-to-average power ratio (PAPR) problems \cite{davis99,paterson,popovic,schmidt}. In 1999, Davis and Jedwab \cite{davis99} proposed $2^h$-ary GCPs ($h\in \mathbb{N}$) based on generalized Boolean functions (GBFs) to control the high PAPR of OFDM systems. The authors proved that the maximum PAPR of a GCP is upper bounded by $2$. Paterson \cite{paterson} further extended the result to $q$-ary (for even $q$) GCPs and proved that the PAPR of CSs can be at most equal to the set size, i.e., the number of sequences in the CS. Since both the constructions were based on GBFs, the constituent sequences in the resultant CSs only have length of the form $2^\alpha$. Although the upper bound of PAPR of a CS is not based on the length of the constituent sequences \cite{paterson}, there are certain practical scenarios where CSs having different lengths other than $2^\alpha$ can be used efficiently. For example, the number of occupied sub-carriers for $20$ MHz bandwidth is $1200$ in long term evolution (LTE) systems where a number of sub-carriers at band edges are reserved for guard bands to prevent the adjacent channel interference \cite{chen}. Recently, in search of CSs having sequences of length other than $2^\alpha$, Chen \cite{chen} in 2016 proposed a novel construction of CSs of length $2^{m-1}+2^v~(m\in \mathbb{N},~0\leq v <m )$ for set size of $4$ and length $2^{m-1}+1$ for set size $2^{k+1}$ by using truncated GBFs. Extending the idea further, Wang \textit{et al.} \cite{wang} in 2017 proposed a new construction of CSs by GBFs which consists of sequences of length $2^{m-1}+2^v$ when the set size is $2^{k+1}$, ($k\in \mathbb{N}$). In 2018, Chen \cite{chen1} proposed some more constructions of CSs with flexible set size and sequence lengths which offers more flexibilities in practical communication systems. Recently, in 2019, Adhikary \textit{et al.} \cite{Avik_cs} proposed CSs of set size $4$ having sequence lengths $M+1$, $M+2$ and CSs of set size $8$ having sequence lengths $2M+3$, where $M$ is is the length of a complementary pair. The construction in \cite{Avik_cs} is based on applying insertion method on Golay pairs. For binary cases, the constituent sequences of the resultant CSs in \cite{Avik_cs},  can sometimes be seen as odd-length binary Z-complementary pairs (OB-ZCPs) \cite{Avik_iwsda,Avik_tit} or even-length binary Z-complementary pairs (EB-ZCPs) \cite{Avik_ebzcp}. Adhikary \textit{et al.} \cite{Avik_cs} closed the paper proposing an open problem of constructing CSs of set size $4$ having sequences of length $29$. 
	
	
	Motivated by the works of Chen \cite{chen,chen1}, Wang \textit{et al.} \cite{wang} and Adhikary \textit{et al.} \cite{Avik_cs}, we propose a generalized construction of CSs of set size $4$ consisting sequences of lengths $M+N$, by concatenating CSs of set size $4$ and lengths $M$ and $N$ , where $M$ and $N$ are lengths for which complementary sequence pairs exists. The proposed construction can easily be extended to a set size of $4n$ (where $n\in\mathbb{N}$). We also propose a new construction of CSs of length $M+P$ and set size $8$, by concatenation of two CSs of set size $8$ and lengths $M$ and $P$, where $M$ is a length for which complementary pair exists and $P$ is a length for which a complementary set of set size $4$ exists. The construction can be easily extended to a set size of $8n$. Up to length $L\leq 34$, for $q=2 \text{ and }4$, our proposed construction can generate CSs given in Table \ref{tab1}. Remember, for $q=2$, the GCPs are available of the form $2^\alpha 10^\beta 26^\gamma$, and for $q=4$, the GCPs are available of the form $ 2^{\alpha+u}3^\beta5^\gamma11^\eta13^\zeta$ where $\alpha,\beta,\gamma,\eta,\zeta,u \geq 0$; $\beta+\gamma+\eta+\zeta\leq \alpha+2u+1$; $u\leq \gamma+\zeta$ \cite{frank,craigen}. The lengths of the CSs of set size 4 and 8, given in bold in Table \ref{tab1}, can not be generated by \cite{chen,chen1,Avik_cs}. We partially solve the open problem proposed by Adhikary \textit{et al.} \cite{Avik_cs} by constructing CSs of set size $4$ having sequence lengths $29$, when $q=4$. This problem for the binary case is still left open. One important contribution of our work is that our constructions generalises all the previously reported constructions along with generating CSs consisting sequences of new lengths. 
	
	\begin{table}[h]
		{
		\centering
		\caption{Proposed non-power-of-two CSs of size 4 and 8, and lengths up to $34$, when $q=2 \text{ and }4$.\label{tab1}}
		\begin{tabular}{ |p{0.9cm}| l | p{5.9cm} |}
			\hline
			& & Lengths \\ \hline
			CSs of set size & $q=2$ & 3, 4, 5, 6, 8, 9, 10, 11, 12, \textbf{14}, 16, 17, 18, 20, 21, 22, \textbf{24}, \textbf{26}, 27, 28, \textbf{30}, 33, 34\\ \cline{2-3}
			4& $q=4$ & 3, 4, 5, 6, 7, 8, 9, 10, 11, 12, 13, 14, 15, 16, 17, 18, 19, 20, 21, 22, 23, 24, 25, 26, 27, 28, \textbf{29}, \textbf{30}, 31, 32, 33, 34 \\ \hline
			CSs of set size 8& $q=2$ & 3, 4, 5, 6, 7, 8, 9, 10, 11, 12, \textbf{13}, \textbf{14}, \textbf{15}, 16,  17, 18, 19, 20, 21, 22, 23, \textbf{24}, \textbf{25}, \textbf{26}, 27, 28, \textbf{29}, \textbf{30}, \textbf{31}, 32, 33, 34 \\ \cline{2-3}
			& $q=4$ & 3, 4, 5, 6, 7, 8, 9, 10, 11, 12, 13, 14, 15, 16, 17, 18, 19, 20, 21, 22, 23, 24, 25, 26, 27, 28, 29, \textbf{30}, 31, 32, 33, 34 \\ \hline
		\end{tabular}}
	\end{table}

The rest of the paper is organised as follows. In Section II, we have defined the aperiodic auto-correlation and insertion function. In Section III, we have proposed constructions of CSs of set size 4 and 8 having different sequences of length. Our proposed constructions are compared with the previous works in Section IV. We conclude our work in Section V.

	\section{Definitions and Notations}
	Let us fix the following notations, which will be used throughout the paper.
	\begin{itemize}
		\item ``$\mathbf{a} || \mathbf{b}$" denotes the horizontal concatenation of sequences $\mathbf{a}$ and $\mathbf{b}$.
		\item $\overleftarrow{\mathbf{a}}$ denotes the reverse of the sequence $\mathbf{a}$.
		\item $\mathbf{a}|_M$ denotes the first $M$ elements of $\mathbf{a}$.
		\item $1$, $-1$, and $-i$ are denoted by $+$, $-$, and $\hat{i}$, respectively.
		\item $\mathbb{C}$ denotes set of all complex numbers. 
		\item $x^*$ denotes the conjugate of an element $x$.
		\item For the notation $[\mathbf{a_0}^T,\dots,\mathbf{a_{n-1}}^T]^T$, it is always assumed that the constituent sequences $\mathbf{a_0},\dots,\mathbf{a_{n-1}}$ are all of same length.
		\item Let $\mathcal{A}\equiv[\mathbf{a}^T,\mathbf{b}^T]^T $. Then $x\mathcal{A}$ means $x$ is multiplied to all the elements of $\mathbf{a}$ and $\mathbf{b}$.
		\item $\mathbb{U}_q=\{\exp{\frac{2\pi\sqrt{-1}t}{q}}:0\leq t <q\}$.
	\end{itemize}

	
	\begin{definition}
		The aperiodic cross-correlation function (ACCF) for two length-$N$ $q$-ary sequences ${\mathbf{a}}$ and ${\mathbf{b}}$ is defined as 
		\begin{equation}\label{defi_ACCF}
		\rho_{\mathbf{a},\mathbf{b}}(\tau):= \left \{
		\begin{array}{cl}
		\sum\limits_{k=0}^{N-1-\tau}{a_kb^*_{k+\tau}},&~~0\leq \tau \leq N-1;\\
		\sum\limits_{k=0}^{N-1-\tau}{a_{k+\tau}b^*_k},&~~-(N-1)\leq \tau \leq -1;\\
		0,& ~~\mid \tau \mid \geq N.
		\end{array}
		\right .
		\end{equation}
		When $\mathbf{a} = \mathbf{b}$, $\rho_{\mathbf{a},\mathbf{b}}(\tau)$ is called aperiodic auto-correlation function (AACF) of $\mathbf{a}$ and is denoted as $\rho_{\mathbf{a}}(\tau)$ \cite{Avik_ebzcp}.
	\end{definition}

	\begin{definition}
		Let $\mathcal{A}=[\mathbf{a}_0^{{T}},~ \mathbf{a}_1^{{T}},~ \dots, \mathbf{a}_{P-1}^{{T}}]^T$ be a $P \times N$ matrix, consisting of $P$ sequences of length $N$. Then $\mathcal{A}$ is called a CS of size $P$ if
		\begin{equation}
		\rho_{\mathbf{a}_0}(\tau)+\rho_{\mathbf{a}_1}(\tau)+\cdots+\rho_{\mathbf{a}_{P-1}}(\tau)=\begin{cases}
		0 & \text{if } 0<\tau<N,\\
		PN & \text{if } \tau=0.
		\end{cases}
		\end{equation}
	\end{definition}
	
%

	\section{Proposed Constructions}
	In this section, we propose two constructions of $q$-ary CSs. In \textit{Theorem \ref{th_const1}}, we give the constructions of CSs having set size $4$. \textit{Theorem 2} constructs CSs of set size $8$. CSs of set size $4n$ and $8n$ can be generated iteratively by vertically concatenating $n$ times. 
	
%
	
	

%

	\begin{theorem}\label{th_const1}
		Consider two $q$-ary complementary pairs $\mathcal{A}\equiv[\mathbf{a}^T,\mathbf{b}^T]^T$ and $\mathcal{B}\equiv[\mathbf{c}^T,\mathbf{d}^T]^T$ of lengths $M$ and $N$, respectively. Let $\mathcal{C}$ be as follows
		\begin{equation}
			\mathcal{C}=\begin{bmatrix}
			x_0\mathcal{A} & y_0\mathcal{B} \\
			x_1\mathcal{A} & y_1\mathcal{B}
			\end{bmatrix},
		\end{equation}
		where $\{x_0,~y_0,~x_1,~y_1\}\in \mathbb{U}_q$. Then $\mathcal{C}$ is a CS of size $4$ and length $M+N$, if the following condition holds
		\begin{equation}
			x_0y_0^*+x_1y_1^*=0.
		\end{equation}
	\end{theorem}


\begin{IEEEproof}
	Let us denote the constituent sequences of $\mathcal{C}$ as $\mathcal{C}=[\mathbf{e}^{{T}},\mathbf{f}^{{T}},\mathbf{g}^{{T}},\mathbf{h}^{{T}}]^T$. Then $\mathbf{e}=x_0\mathbf{a}\mid \mid y_0\mathbf{c}$, $\mathbf{f}=x_0\mathbf{b}\mid \mid y_0\mathbf{d}$, $\mathbf{g}=x_1\mathbf{a}\mid \mid y_1\mathbf{c}$ and $\mathbf{h}=x_1\mathbf{b}\mid \mid y_1\mathbf{d}$.
	
Let $M<N$, then for $1\leq \tau \leq M-1$ we have,
\begin{equation}
\begin{split}
	\rho_{\mathbf{e}}(\tau)+\rho_{\mathbf{g}}(\tau)=&(x_0x_0^*+x_1x_1^*)\rho_{\mathbf{a}}(\tau)+(y_0y_0^*+y_1y_1^*)\rho_{\mathbf{c}}(\tau)\\&\hspace{.5cm}+(x_0y_0^*+x_1y_1^*)\rho_{\overleftarrow{\mathbf{a}},\overleftarrow{\mathbf{c}|_{M}}}(M-\tau),
\end{split}
\end{equation}
	for $M\leq \tau \leq N-1$ we have,
	\begin{equation}
	\begin{split}
	\rho_{\mathbf{e}}(\tau)+\rho_{\mathbf{g}}(\tau)=&(x_0y_0^*+x_1y_1^*)\rho_{\mathbf{a},\mathbf{c}|_M}(\tau-M)\\&\hspace{1cm}+(y_0y_0^*+y_1y_1^*)\rho_{\mathbf{c}}(\tau),
	\end{split}
	\end{equation}
	for $N\leq \tau < M+N$ we have,
	\begin{equation}
	\rho_{\mathbf{e}}(\tau)+\rho_{\mathbf{g}}(\tau)=(x_0y_0^*+x_1y_1^*)\rho_{\mathbf{a},\overleftarrow{\mathbf{c}}|_M}(\tau-N).
	\end{equation}	
	By similar calculations we have for $1\leq \tau \leq M-1$ we have,
\begin{equation}
\begin{split}
\rho_{\mathbf{f}}(\tau)+\rho_{\mathbf{h}}(\tau)=&(x_0x_0^*+x_1x_1^*)\rho_{\mathbf{b}}(\tau)+(y_0y_0^*+y_1y_1^*)\rho_{\mathbf{d}}(\tau)\\&\hspace{.5cm}+(x_0y_0^*+x_1y_1^*)\rho_{\overleftarrow{\mathbf{b}},\overleftarrow{\mathbf{d}|_{M}}}(M-\tau),
\end{split}
\end{equation}
for $M\leq \tau \leq N-1$ we have,
\begin{equation}
\begin{split}
\rho_{\mathbf{f}}(\tau)+\rho_{\mathbf{h}}(\tau)=&(x_0y_0^*+x_1y_1^*)\rho_{\mathbf{b},\mathbf{d}|_M}(\tau-M)\\&\hspace{1cm}+(y_0y_0^*+y_1y_1^*)\rho_{\mathbf{d}}(\tau),
\end{split}
\end{equation}
for $N\leq \tau < M+N$ we have,
\begin{equation}
\rho_{\mathbf{f}}(\tau)+\rho_{\mathbf{h}}(\tau)=(x_0y_0^*+x_1y_1^*)\rho_{\mathbf{b},\overleftarrow{\mathbf{d}}|_M}(\tau-N).
\end{equation}	
Therefore, for $0<\tau<M+N$,
\begin{equation}
	\rho_{\mathbf{e}}(\tau)+\rho_{\mathbf{g}}(\tau)+\rho_{\mathbf{f}}(\tau)+\rho_{\mathbf{h}}(\tau)=0,
\end{equation}
if and only if
\begin{equation}
	x_0y_0^*+x_1y_1^*=0.
\end{equation}

 Similarly, we can also prove the theorem for the cases when $M=N$ or $M>N$. This completes the proof.
\end{IEEEproof}


\begin{example}\label{ex1}
	Let us consider a GCP $(\mathbf{a},\mathbf{b})$ of length $10$ as follows:
	\begin{equation}
	\begin{array}{lcl}
	\mathbf{a}& =& (++--+++-+-),\\
	\mathbf{b}& =& (+++++-+--+).
	\end{array}
	\end{equation}
	and $(\mathbf{c},\mathbf{d})$ be a GCP of length $4$ as follows,
	\begin{equation}
	\begin{array}{lcl}
	\mathbf{c}& =& (++-+),\\
	\mathbf{d}& =& (+++-).
	\end{array}
	\end{equation}
	Let $x_0=1,~x_1=1,~y_0=1$ and $y_1=-1$. Then according to Theorem \ref{th_const1}, we have 
	\begin{equation}
	\mathcal{C}=\begin{bmatrix}
	\mathcal{A} & \mathcal{B} \\
	\mathcal{A} & -\mathcal{B}
	\end{bmatrix},
	\end{equation}
	where $\mathcal{A}\equiv[\mathbf{a}^T,\mathbf{b}^T]^T$ and $\mathcal{B}\equiv[\mathbf{c}^T,\mathbf{d}^T]^T$. Then $\mathbf{e}=\mathbf{a}\mid \mid \mathbf{c}$, $\mathbf{f}=\mathbf{b}\mid \mid \mathbf{d}$, $\mathbf{g}=\mathbf{a}\mid \mid -\mathbf{c}$ and $\mathbf{h}=\mathbf{b}\mid \mid -\mathbf{d}$, are given as,
	\begin{equation}
	\begin{array}{lcl}
	\mathbf{e}& =& ({\color{red}++--+++-+-}{\color{blue}++-+} ),\\
	\mathbf{f}& =& ({\color{red}+++++-+--+}{\color{blue}+++-}),\\
	\mathbf{g}& =& ({\color{red}++--+++-+-}{\color{blue}--+-}),\\
	\mathbf{h}& =& ({\color{red}+++++-+--+}{\color{blue}---+}).
	\end{array}
	\end{equation}
	$\mathcal{C}=[\mathbf{e}^{{T}},\mathbf{f}^{{T}},\mathbf{g}^{{T}},\mathbf{h}^{{T}}]^T$ is a CS of set size $4$ consisting sequences of length $14$. Please note that for $q=2$, CSs of set size $4$, consisting sequences of length $14$ have not been reported before.
\end{example}

\begin{remark}
	In the above example, for $q=4$, if we use $(\mathbf{a},\mathbf{b})$ as a complementary pair of length $3$ and $(\mathbf{c},\mathbf{d})$ as a complementary pair of length $26$, then we can generate a CS of set size $4$ having length $29$. This partially answers the open problem proposed in \cite{Avik_cs}.
\end{remark}

\begin{remark}
	Exhaustive computer search suggests that for $q=4$, the proposed construction in Theorem \ref{th_const1} can generate CSs of set size $4$ consisting sequences of any length up to $86$. For $q=4$, the problem of generating CSs of set size $4$ consisting sequences of length $87$ is still open.
\end{remark}

\begin{theorem}\label{th2}
	Consider a $q$-ary GCP $(\mathbf{a},\mathbf{b})$ of length $M$, and a $q$-ary complementary set $\mathcal{B}$ of size $4$ having constituent sequences of length $P$. Let $\mathcal{C}$ be denoted as follows 
	\begin{equation}
		\mathcal{C}=\begin{bmatrix}
		\begin{matrix}
		x_0\mathcal{A}\\
		x_1\mathcal{A}
		\end{matrix} & y_0\mathcal{B}\\
		\begin{matrix}            
		x_2\mathcal{A}\\ 
		x_3\mathcal{A}
		\end{matrix}& y_1\mathcal{B}\\             
		\end{bmatrix},
	\end{equation}
	where $\{x_0,~x_1,~x_2,~x_3,~y_0,~y_1\}\in \mathbb{U}_q$. Then $\mathcal{C}$ is a CS of set size $8$ having constituent sequences of length $M+P$, if the following conditions hold:
	\begin{equation}
	\begin{split}
	x_0y_0^*+x_2y_1^*&=0 \text{ and }\\
	x_1y_0^*+x_3y_1^*&=0.
	\end{split}
	\end{equation}
\end{theorem}
\begin{IEEEproof}
	
Let us consider $\mathcal{B}=[\mathbf{e}^{{T}},\mathbf{f}^{{T}},\mathbf{g}^{{T}},\mathbf{h}^{{T}}]^T$, a CS of size $4$ and constituent sequences of length $P$. Also, let $\mathbf{p_0}=x_0\mathbf{a}\mid \mid y_0\mathbf{e}$, $\mathbf{p_1}=x_0\mathbf{b}\mid \mid y_0\mathbf{f}$, $\mathbf{p_2}=x_1\mathbf{a}\mid \mid y_0\mathbf{g}$,
$\mathbf{p_3}=x_1\mathbf{b}\mid \mid y_0\mathbf{h}$,
$\mathbf{p_4}=x_2\mathbf{a}\mid \mid y_1\mathbf{e}$,
$\mathbf{p_5}=x_2\mathbf{b}\mid \mid y_1\mathbf{f}$, $\mathbf{p_6}=x_3\mathbf{a}\mid \mid y_1\mathbf{g}$ and
$\mathbf{p_7}=x_3\mathbf{b}\mid \mid y_1\mathbf{h}$. We will prove the theorem assuming $M<P$. The other cases can be proved similarly. For $1\leq \tau \leq M-1$, we have
\begin{equation}
\begin{split}
\rho_{\mathbf{p_0}}(\tau)&=x_0x_0^*\rho_{\mathbf{a}}(\tau)+x_0y_0^*\rho_{\overleftarrow{\mathbf{a}},\overleftarrow{\mathbf{e}|_{M}}}(M-\tau)+y_0y_0^*\rho_{\mathbf{e}}(\tau),\\
\rho_{\mathbf{p_1}}(\tau)&=x_0x_0^*\rho_{\mathbf{b}}(\tau)+x_0y_0^*\rho_{\overleftarrow{\mathbf{b}},\overleftarrow{\mathbf{f}|_{M}}}(M-\tau)+y_0y_0^*\rho_{\mathbf{f}}(\tau),\\
\rho_{\mathbf{p_2}}(\tau)&=x_1x_1^*\rho_{\mathbf{a}}(\tau)+x_1y_0^*\rho_{\overleftarrow{\mathbf{a}},\overleftarrow{\mathbf{g}|_{M}}}(M-\tau)+y_0y_0^*\rho_{\mathbf{g}}(\tau),\\
\rho_{\mathbf{p_3}}(\tau)&=x_1x_1^*\rho_{\mathbf{b}}(\tau)+x_1y_0^*\rho_{\overleftarrow{\mathbf{b}},\overleftarrow{\mathbf{h}|_{M}}}(M-\tau)+y_0y_0^*\rho_{\mathbf{h}}(\tau),\\
\rho_{\mathbf{p_4}}(\tau)&=x_2x_2^*\rho_{\mathbf{a}}(\tau)+x_2y_1^*\rho_{\overleftarrow{\mathbf{a}},\overleftarrow{\mathbf{e}|_{M}}}(M-\tau)+y_1y_1^*\rho_{\mathbf{e}}(\tau),\\
\rho_{\mathbf{p_5}}(\tau)&=x_2x_2^*\rho_{\mathbf{b}}(\tau)+x_2y_1^*\rho_{\overleftarrow{\mathbf{b}},\overleftarrow{\mathbf{f}|_{M}}}(M-\tau)+y_1y_1^*\rho_{\mathbf{f}}(\tau),\\
\rho_{\mathbf{p_6}}(\tau)&=x_3x_3^*\rho_{\mathbf{a}}(\tau)+x_3y_1^*\rho_{\overleftarrow{\mathbf{a}},\overleftarrow{\mathbf{g}|_{M}}}(M-\tau)+y_1y_1^*\rho_{\mathbf{g}}(\tau),\\
\rho_{\mathbf{p_7}}(\tau)&=x_3x_3^*\rho_{\mathbf{b}}(\tau)+x_3y_1^*\rho_{\overleftarrow{\mathbf{b}},\overleftarrow{\mathbf{h}|_{M}}}(M-\tau)+y_1y_1^*\rho_{\mathbf{h}}(\tau).
\end{split}
\end{equation}
For $M\leq \tau \leq P-1$, we have	
\begin{equation}
\begin{split}
	\rho_{\mathbf{p_0}}(\tau)&=x_0y_0^*\rho_{\mathbf{a},\mathbf{e}|_M}(\tau-M)+y_0y_0^*\rho_{\mathbf{e}}(\tau),\\
	\rho_{\mathbf{p_1}}(\tau)&=x_0y_0^*\rho_{\mathbf{b},\mathbf{f}|_M}(\tau-M)+y_0y_0^*\rho_{\mathbf{f}}(\tau),\\
	\rho_{\mathbf{p_2}}(\tau)&=x_1y_0^*\rho_{\mathbf{a},\mathbf{g}|_M}(\tau-M)+y_0y_0^*\rho_{\mathbf{g}}(\tau),\\
	\rho_{\mathbf{p_3}}(\tau)&=x_1y_0^*\rho_{\mathbf{b},\mathbf{h}|_M}(\tau-M)+y_0y_0^*\rho_{\mathbf{h}}(\tau),\\
	\rho_{\mathbf{p_4}}(\tau)&=x_2y_1^*\rho_{\mathbf{a},\mathbf{e}|_M}(\tau-M)+y_1y_1^*\rho_{\mathbf{e}}(\tau),\\
	\rho_{\mathbf{p_5}}(\tau)&=x_2y_1^*\rho_{\mathbf{b},\mathbf{f}|_M}(\tau-M)+y_1y_1^*\rho_{\mathbf{f}}(\tau),\\
	\rho_{\mathbf{p_6}}(\tau)&=x_3y_1^*\rho_{\mathbf{a},\mathbf{g}|_M}(\tau-M)+y_1y_1^*\rho_{\mathbf{g}}(\tau),\\
	\rho_{\mathbf{p_7}}(\tau)&=x_3y_1^*\rho_{\mathbf{b},\mathbf{h}|_M}(\tau-M)+y_1y_1^*\rho_{\mathbf{h}}(\tau).
\end{split}
\end{equation}	
For $P\leq \tau < M+P$, we have	
\begin{equation}
\begin{split}
	\rho_{\mathbf{p_0}}(\tau)&=x_0y_0^*\rho_{\mathbf{a},\overleftarrow{\mathbf{e}}|_M}(\tau-P),\\
	\rho_{\mathbf{p_1}}(\tau)&=x_0y_0^*\rho_{\mathbf{b},\overleftarrow{\mathbf{f}}|_M}(\tau-P),\\
	\rho_{\mathbf{p_2}}(\tau)&=x_1y_0^*\rho_{\mathbf{a},\overleftarrow{\mathbf{g}}|_M}(\tau-P),\\
	\rho_{\mathbf{p_3}}(\tau)&=x_1y_0^*\rho_{\mathbf{b},\overleftarrow{\mathbf{h}}|_M}(\tau-P),\\
	\rho_{\mathbf{p_4}}(\tau)&=x_2y_1^*\rho_{\mathbf{a},\overleftarrow{\mathbf{e}}|_M}(\tau-P),\\
	\rho_{\mathbf{p_5}}(\tau)&=x_2y_1^*\rho_{\mathbf{b},\overleftarrow{\mathbf{f}}|_M}(\tau-P),\\
	\rho_{\mathbf{p_6}}(\tau)&=x_3y_1^*\rho_{\mathbf{a},\overleftarrow{\mathbf{g}}|_M}(\tau-P),\\
	\rho_{\mathbf{p_7}}(\tau)&=x_3y_1^*\rho_{\mathbf{b},\overleftarrow{\mathbf{h}}|_M}(\tau-P),\\
\end{split}
\end{equation}	
Therefore, for $1\leq \tau \leq M-1$,
\begin{equation}
	\sum_{i=0}^{7}\rho_{\mathbf{p_i}}(\tau)=0,
\end{equation}
if and only if
\begin{equation}\label{eq22}
\begin{split}
&(x_0y_0^*+x_2y_1^*)(\rho_{\overleftarrow{\mathbf{a}},\overleftarrow{\mathbf{e}|_{M}}}(M-\tau)+\rho_{\overleftarrow{\mathbf{b}},\overleftarrow{\mathbf{f}|_{M}}}(M-\tau))+\\&(x_1y_0^*+x_3y_1^*)(\rho_{\overleftarrow{\mathbf{a}},\overleftarrow{\mathbf{g}|_{M}}}(M-\tau)+\rho_{\overleftarrow{\mathbf{b}},\overleftarrow{\mathbf{h}|_{M}}}(M-\tau))=0.
\end{split}
\end{equation}
Eq. (\ref{eq22}) is zero when
\begin{equation}
\begin{split}
	x_0y_0^*+x_2y_1^*&=0 \text{ and }\\
    x_1y_0^*+x_3y_1^*&=0.
\end{split}
\end{equation}
For $M\leq \tau \leq P-1$,
\begin{equation}
\sum_{i=0}^{7}\rho_{\mathbf{p_i}}(\tau)=0,
\end{equation}
if and only if
\begin{equation}\label{eq25}
\begin{split}
	&(x_0y_0^*+x_2y_1^*)(\rho_{\mathbf{a},\mathbf{e}|_M}(\tau-M)+\rho_{\mathbf{b},\mathbf{f}|_M}(\tau-M))+\\
&(x_1y_0^*+x_3y_1^*)(\rho_{\mathbf{a},\mathbf{g}|_M}(\tau-M)+\rho_{\mathbf{b},\mathbf{h}|_M}(\tau-M))=0.
\end{split}
\end{equation}
Eq. (\ref{eq25}) is zero when
\begin{equation}
	\begin{split}
	x_0y_0^*+x_2y_1^*&=0 \text{ and }\\
	x_1y_0^*+x_3y_1^*&=0.
	\end{split}
\end{equation}
Similarly, for $P\leq \tau <M+P$
\begin{equation}
\sum_{i=0}^{7}\rho_{\mathbf{p_i}}(\tau)=0,
\end{equation}
when
\begin{equation}
\begin{split}
x_0y_0^*+x_2y_1^*&=0 \text{ and }\\
x_1y_0^*+x_3y_1^*&=0.
\end{split}
\end{equation}	
This completes the proof.
Similarly, we can prove the above theorem for $M=P$ or $M>P$.
\end{IEEEproof}

\begin{example}\label{ex2}
	Let us consider a complementary pair $(\mathbf{a},\mathbf{b})$ of length $8$ as follows:
	\begin{equation}
	\begin{array}{lcl}
	\mathbf{a}& =& (+++-++-+),\\
	\mathbf{b}& =& (+++---+-).
	\end{array}
	\end{equation}
	and $\mathcal{B}=[\mathbf{e}^{{T}},\mathbf{f}^{{T}},\mathbf{g}^{{T}},\mathbf{h}^{{T}}]^T$ be a CS of set size $4$ consisting sequences of length $5$, as follows,
	\begin{equation}
	\mathcal{B}=\begin{bmatrix}
	++-++ \\
	+++-- \\
	+-+++ \\
	+-+--
	\end{bmatrix},
	\end{equation}
	Let $x_0=1,~x_1=-1,~x_2=-1,~x_3=1,~y_0=1$ and $y_1=1$. Then according to Theorem \ref{th2}, we have 
	\begin{equation}
	\mathcal{C}=\begin{bmatrix}
	\begin{matrix}
	\mathcal{A}\\
	-\mathcal{A}
	\end{matrix} & \mathcal{B}\\
	\begin{matrix}            
	-\mathcal{A}\\ 
	\mathcal{A}
	\end{matrix}& \mathcal{B}\\             
	\end{bmatrix},
	\end{equation}
	where $\mathcal{A}\equiv[\mathbf{a}^T,\mathbf{b}^T]^T$. Then $\mathbf{p_0}=\mathbf{a}\mid \mid \mathbf{e}$, $\mathbf{p_1}=\mathbf{b}\mid \mid \mathbf{f}$, $\mathbf{p_2}=-\mathbf{a}\mid \mid \mathbf{g}$,
	$\mathbf{p_3}=-\mathbf{b}\mid \mid \mathbf{h}$,
	$\mathbf{p_4}=-\mathbf{a}\mid \mid \mathbf{e}$,
	$\mathbf{p_5}=-\mathbf{b}\mid \mid \mathbf{f}$, $\mathbf{p_6}=\mathbf{a}\mid \mid \mathbf{g}$ and
	$\mathbf{p_7}=\mathbf{b}\mid \mid \mathbf{h}$, are given as,
	\begin{equation}
	\begin{array}{lcl}
	\mathbf{p_0}& =& ({\color{red}+++-++-+}{\color{blue}++-++} ),\\
	\mathbf{p_1}& =& ({\color{red}+++---+-}{\color{blue}+++--}),\\
	\mathbf{p_2}& =& ({\color{red}---+--+-}{\color{blue}+-+++}),\\
	\mathbf{p_3}& =& ({\color{red}---+++-+}{\color{blue}+-+--}),\\
	\mathbf{p_4}& =& ({\color{red}---+--+-}{\color{blue}++-++} ),\\
	\mathbf{p_5}& =& ({\color{red}---+++-+}{\color{blue}+++--}),\\
	\mathbf{p_6}& =& ({\color{red}+++-++-+}{\color{blue}+-+++}),\\
	\mathbf{p_7}& =& ({\color{red}+++---+-}{\color{blue}+-+--}).
	\end{array}
	\end{equation}
	$\mathcal{C}=[\{\mathbf{p_i}^T\}_{i=0}^7]^T$ is a CS of set size $8$ consisting sequences of length $13$. Please note that for $q=2$, CSs of set size $8$, consisting sequences of length $13$ have not been reported before.
\end{example}


\section{Comparison With the Previous Works}
The proposed construction generalises the previous constructions in terms of the length of the constituent sequences in the CS. The proposed construction also generates CSs with the constituent sequences having new lengths which are not reported before. Table \ref{compare1} and Table \ref{compare2} contain a detail comparison with the previous works.

\begin{table}[h]
	\caption{Parameters of CSs of Set Size 4.\label{compare1}}
	\begin{tabular}{|c|c|c|c|}
		\hline
		Ref.  & Length & Constraints \\ \hline
		\cite{chen} & $2^{m-1}+2^v$ & $0\leq v \leq m-1$ \\ \hline
		\cite{chen1} &   $2^{m-1}+\alpha2^{\pi(m-1)-1}+2^v$  & \makecell{$\alpha\in\{0,1\}$,\\$0\leq v \leq m-2$}  \\ \hline
		\cite{chen2017} & $2^{m-1}+2^v$ & $m\leq 2,1\leq v< m-1$    \\ \hline
		\cite{wang}&$2^{m-1}+2^v$&$m\leq 3,2\leq v\leq m-1$    \\ \hline
		\cite{Avik_cs}&$N+1,~N+2$&\makecell{$N$ is the length of\\ complementary pairs}   \\ \hline
		Proposed  & $M+N$ & \makecell{$M$,$N$ are the lengths of\\  complementary sequence\\ pairs}  \\ \hline
	\end{tabular}
\end{table}

\begin{table}[h]
	\caption{Parameters of CSs of Set Size 8.\label{compare2}}
	\begin{tabular}{|c|c|c|c|}
		\hline
		Ref.  & Length & Constraints \\ \hline
		\cite{chen} & $2^{m-1}+2^v$ & $0\leq v \leq m-1$ \\ \hline
		\cite{chen1} &   $2^{m-1}+\alpha2^{\pi(m-1)-1}+2^v$  & \makecell{$\alpha\in\{0,1\}$,\\$0\leq v \leq m-2$}  \\ \hline
		\cite{chen2017} & $2^{m-1}+2^v$ & $m\leq 2,1\leq v< m-1$    \\ \hline
		\cite{wang}&$2^{m-1}+2^v$&$m\leq 3,2\leq v\leq m-1$    \\ \hline
		\cite{Avik_cs}&$N+1,~N+2,~2N+3$&\makecell{$N$ is the length of\\ complementary pairs}   \\ \hline
		Proposed  & $M+N,~M+P$ & \makecell{$M$,$N$ are the lengths of\\  complementary sequence \\pairs and $P$ is the length \\of CSs of set size $4$.}  \\ \hline
	\end{tabular}
\end{table}

\section{Conclusion}
In this paper, we have given new constructions of $q$-ary CSs of length $M+N$ of size $4$, where $M$ and $N$ are lengths of two GCPs. We have also constructed $q$-ary CSs of length $M+P$ and set size $8$, where $M$ is the length of a GCP and $P$ is the length of a constituent sequence of a size $4$ CS. Moreover, the CSs can be generated iteratively, and concatenated vertically to increase the set size to $4n$ or $8n$, where $n$ is precisely the number of iterations. The CSs are constructed by concatenating GCPs maintaining certain conditions. Up to length $L\leq 34$, for $q=2$ and $4$, \textit{Theorem \ref{th_const1}} proposed systematic construction of CSs of set size $4$, given in Table \ref{tab1}. Recursive application of \textit{Theorem \ref{th_const1}}, \textit{Theorem \ref{th2}}, proposed systematic construction of CSs of set size $8$, also given in Table \ref{tab1}. 
These constructions partially fulfils the gaps left by the constructions proposed previously by Chen and Adhikary \textit{et al.}. A possible future research is to construct CSs of other lengths which are not reported till date, like binary CSs of set size 4 and length 29.



	\end{document}